\documentclass[11pt]{article}

\usepackage{amsfonts,amssymb,amsmath,amsthm,amscd,latexsym, epsfig}

\def\be{\begin{equation}}
\def\ee{\end{equation}}
\def\bea{\begin{eqnarray}}
\def\eea{\end{eqnarray}}

\def\1{\'{\i}}                           
\def\rsl{\mathfrak{sl}} 
\def\RR{\mathbb{R}} 

\def\m{{\eta}}  
\def\pairing{{\cal P}} 
\def\calD{{\cal D}} 
 
 \def\lam{{\lambda}}
\newcommand{\oo}[0]{\otimes}

\begin{document}

\begin{center}
\baselineskip 24 pt {\Large \bf  
Three-dimensional  gravity and Drinfel'd doubles:\\ spacetimes and symmetries from quantum  deformations}



\end{center}

\medskip

\begin{center}   {\sc Angel~Ballesteros$^a$,
 Francisco~J.~Herranz$^a$ and Catherine Meusburger$^{b}$ }
\end{center}

\begin{center} {\it { 
${}^a$Departamento de F\1sica, University of Burgos, E-09001 Burgos, Spain }}\\ E-mail:
angelb@ubu.es, fjherranz@ubu.es
\end{center}

\begin{center} {\it { 
${}^b$ Department Mathematik,
Universit\"at Hamburg,
Bundestra\ss e 55, 
D-20146 Hamburg,
Germany}}\\  
E-mail:  catherine.meusburger@uni-hamburg.de
\end{center}

\medskip

\begin{abstract}

We show how the constant curvature spacetimes of 3d gravity and the associated symmetry algebras can be derived from a single  quantum deformation of the 3d Lorentz algebra $\rsl(2,\RR)$. 
We investigate the classical Drinfel'd double of  a ``hybrid"  deformation of $\rsl(2,\RR)$ that depends on two parameters $(\m,z)$. With an appropriate choice of basis and real structure, this Drinfel'd double 
agrees with the 3d anti-de Sitter algebra. The deformation parameter  $\m$ is related to the
cosmological constant, while $z$  is identified with the inverse of the  speed of light  and defines the signature 
of the metric. We generalise this result to de Sitter space, the three-sphere and 3d hyperbolic space through analytic continuation in $\m$ and $z$;  we also investigate the limits of vanishing $\m$ and $z$, which yield the flat spacetimes (Minkowski and Euclidean spaces) and Newtonian models, respectively.

\end{abstract}
 

\medskip 








\section{Introduction}

Quantum group symmetries have been discussed extensively as possible symmetries of a quantum theory of gravity. It is widely believed, see~\cite{amel} and references therein, that the low energy limit of a quantum theory of gravity  would be invariant under certain quantum deformations of the Poincar\'e group. This gave rise  to the so-called  ``doubly special relativity"   
theories~\cite{amel,Amelino-Camelia:2000mrr,Bruno:primo, Amelino-Camelia:2000mn,
MagueijoSmolin,Kowalski-Glikman:2002we,Lukierski:2002df,KowalskiFS},  in which the 
 deformation parameter is interpreted as an
observer-independent  fundamental scale related to the Planck length.  On the other hand, there is a number of models of quantum gravity based on $q$-deformed universal enveloping algebras, in which the deformation parameter is identified with the cosmological constant~\cite{amel,plbcurvature, IJTP,cm1}.  

Of particular interest in this context is 3d gravity, which can be quantised rigourously and in which quantum group symmetries appear naturally as the quantum counterparts of Poisson--Lie symmetries in the classical theory \cite{FR, AMII}. In this case, there is strong evidence \cite{BNR,we2,cm2} that the relevant quantum groups are certain Drinfel'd doubles associated with the isometry groups of Lorentzian and Euclidean constant curvature spacetimes. In fact, the Drinfel'd double approach to obtain deformed spacetime symmetries   was early introduced   in~\cite{Woronowicz}  and applied to the construction of   a one-parameter  quantum Lorentz group.
Moreover, the 3d spacetimes    arising for different signatures 
and   values of the cosmological constant exhibit strong similarities both in their geometrical features \cite{bb} and with respect to their Poisson--Lie and quantum group symmetries \cite{cm1}.  This makes 3d gravity an ideal model for the investigation of the role of deformation parameters and their physical interpretation. Specifically, it allows one to study their role in the physically relevant limits (vanishing cosmological constant, classical limit, vanishing gravitational constant)~\cite{bernd}.

It is therefore natural to search for the underlying mathematical structures which account for these  similarities and to develop a framework that relates the physical parameters of the models to quantum deformations. 
 In this letter we provide a preliminary answer to this question. We show that the constant curvature spacetime in 3d gravity, their isometry groups and the   associated quantum groups all arise from a single quantum deformation of the 3d Lorentz algebra $\rsl(2,\RR)$. Moreover, this quantum deformation supplies the additional structures (star structure and pairing) that enter in the Chern--Simons  formulation of the theory.
This establishes a direct link between quantum deformations of $\rsl(2,\RR)$ and 3d gravity models in which the different  physical limits arise as Lie algebra contractions.

  While most quantum deformations investigated in the context of quantum gravity are based on a {\em single} deformation parameter, 
  we show in this letter that  multi-parametric ones  provide a common framework for 3d gravity. More specifically, we  
 consider  a {\em  two-parametric} quantum deformation of   the   $\rsl(2,\RR)$  Lie algebra with real deformation parameters $\m$ and $z$. This  deformation has a ``hybrid" character since it can be understood as a superposition of the standard (or Drinfel'd--Jimbo) deformation (with parameter $\m$) and  the nonstandard (or Jordanian twist) one (with parameter $z$). We show that both parameters have a direct physical interpretation in the context of 3d gravity: $\m$ corresponds to the cosmological constant, while $z$ is related to the speed of light.

The  letter is structured as follows. In the next section, we give a brief summary of 3d gravity with an emphasis on its formulation as a Chern--Simons   gauge theory.
 In Section 3, we recall the two fundamental  quantum deformations of $\rsl(2,\RR)$ and their role as kinematical symmetries in Planck scale constructions. In Section 4 we construct the classical Drinfel'd double  of the  ``hybrid"  two-parametric 
deformation of $\rsl(2,\RR)$ following~\cite{gomez,ab1}. We show that this double has a natural interpretation as the isometry algebra of 3d anti-de Sitter (AdS) space  with its two parameters corresponding to the cosmological constant and the speed of light. 
In Section 5, we extend our model through
  analytic continuation  in both parameters $\m$ and $z$ and investigate the limits $\m\rightarrow 0$, $z\rightarrow 0$. This yields a unified description
   of nine homogeneous spaces which contains the six constant curvature ones arising in 3d gravity:  the three-sphere, 3d hyperbolic and Euclidean space for Euclidean signature, together with  the 3d AdS, de Sitter (dS) and Minkowski space for Lorentzian signature. The remaining three cases
   correspond to Newtonian (non-relativistic) limits~\cite{IJTP,BLL,Aldrovandi}.  Finally, we comment on our results and present perspectives for   future work.


\section{ Gravity in three dimensions}

The distinguishing feature of 3d gravity  is that the theory has no local gravitational degrees of freedom. Any solution of the 3d vacuum Einstein equations is of constant curvature, which is given by the cosmological constant $\Lambda$, and is locally isometric to one of six standard spacetimes.  For Euclidean signature these are the three-sphere ${\bf S}^3$ ($\Lambda>0$), 3d hyperbolic space ${\bf H}^3$ ($\Lambda<0$) and 3d Euclidean space ${\bf E^3}$ ($\Lambda=0$). For Lorentzian signature, we have  3d dS space ${\bf dS}^{2+1}$ ($\Lambda>0$),   AdS space ${\bf AdS}^{2+1}$ ($\Lambda<0$) and   Minkowski space ${\bf M}^{2+1}$ ($\Lambda=0$). All of these spacetimes are homogeneous spaces and given as quotients of their isometry group by either  the 3d rotation group $SO(3)$ (Euclidean) or Lorentz group $SO(2,1)$ (Lorentzian).


\begin{table}[t] {\footnotesize

 \noindent
\caption{{\small Constant curvature spacetimes and isometry groups  in 3d gravity.
 }}
\vspace{2ex}
\label{sptimes}
  \begin{tabular}{llll}
  \hline
&  $\Lambda>0$ & $\Lambda=0$ & $\Lambda<0$\\ \hline
& & &\\[-2ex]
{\footnotesize Lorentzian} & ${\bf dS}^{2+1}=SO(3,1)/SO(2,1)$ & ${\bf M}^{2+1}=ISO(2,1)/SO(2,1)$ & ${\bf AdS}^{2+1}=SO(2,2)/SO(2,1)$\\
& $\text{Isom}({\bf dS}^{2+1})=SO(3,1)$ & $\text{Isom}({\bf M}^{2+1})=ISO(2,1)$ & $\text{Isom}({\bf AdS}^{2+1})=SO(2,2)$\\[2pt]\hline
& & &\\[-2ex]
{\footnotesize Euclidean}  & ${\bf S}^3= SO(4)/SO(3)$   & ${\bf E}^3=ISO(3)/SO(3)$ & ${\bf H}^3=SO(3,1)/SO(3)$\\
&$\text{Isom}({\bf S}^{3})=SO(4)$   & $\text{Isom}({\bf E}^{3})=ISO(3)$ & $\text{Isom}({\bf H}^{3})=SO(3,1)$\\[2pt] \hline
   \end{tabular}
} 
\end{table}


The absence of local gravitational degrees of freedom in 3d gravity allows one to formulate the theory as a Chern--Simons (CS) gauge theory \cite{AT, Witten1}, where the gauge group is the isometry group of the associated standard spacetime in Table \ref{sptimes} or a cover thereof.
It is shown in \cite{Witten1} that the  Lie algebras of these isometry groups
can be parametrised in terms of  generators $J_a, P_a$, $a=0,1,2$, such that  the cosmological constant and signature arise as
 parameters in the Lie bracket. We have 
  \begin{align}
  \label{brack}
  [J_a,J_b]=\epsilon_{abc}J^c , \qquad [J_a,P_b]=\epsilon_{abc}P^c , \qquad [P_a,P_b]=\lam\epsilon_{abc}J^c,
  \end{align}  
where, depending on the signature, indices are raised with either the 3d Minkowski metric or the 3d Euclidean metric and $\lambda$ is directly related to the cosmological constant $\Lambda$:
\begin{align}\label{lambdadef}
\lam=\begin{cases}\Lambda & \text{for Euclidean signature};\\
-\Lambda  & \text{for Lorentzian signature.}\end{cases}
\end{align}
It is shown in \cite{cm1} that this parametrisation  of the   symmetry algebras gives rise to a unified description of the isometry groups for different signatures and curvature in terms of (pseudo)quaternions over commutative rings.

The CS formulation of 3d gravity is obtained from Cartan's formulation of the theory  by combining the triad $e$ and the spin connection $\omega$ into a CS gauge field. Locally, the gauge field is a one-form on a three-manifold $M$ with values in the Lie algebra \eqref{brack}. 
In units where the speed of light is set to one, it is given by  $$A=e^aP_a+\omega^aJ_a.$$
In order to reproduce the reality conditions of 3d gravity, namely that the triad $e$ and spin connection $\omega$ are real-valued, 
the Lie algebras \eqref{brack} have to be regarded as {\em real} Lie algebras, i.e.~equipped with the star structure
  \begin{align}\label{starcs}
  J_a^*=-J_a,\qquad P_a^*=-P_a.
  \end{align}
In addition to the choice of a Lie algebra, the formulation of a CS gauge theory requires the choice of a symmetric, non-degenerate, Ad-invariant bilinear form $\langle\,,\,\rangle$ on this Lie algebra.  For the Lie algebras \eqref{brack}, the space of symmetric, Ad-invariant bilinear forms is two-dimensional. It is shown in \cite{Witten1}, for a detailed discussion see also \cite{cm2,etera},  that the form relevant for the CS formulation of 3d gravity is given by
   \begin{align}
  \label{pair}
  \langle J_a,P_b\rangle=g_{ab},\qquad \langle J_a,J_b\rangle=\langle P_a,P_b\rangle=0,
  \end{align}
  where, depending on the signature, $g_{ab}$ is either the Euclidean or the Minkowski metric. 
With these choices, the CS action
 \begin{align}
  \label{csact}
  I_{CS}[A]=\int_M \langle A\wedge dA+\frac 1 3 A\wedge[A,A]\rangle,  \end{align}
can be rewritten as the Einstein--Hilbert action for 3d gravity, and the equations of motion derived from it are equivalent to the Einstein equations, namely the requirements of vanishing torsion and constant curvature \cite{Witten1}.

The CS formulation of 3d gravity gave rise to important progress in the description of the phase space and in the quantisation of 3d gravity.  In particular, it is shown in \cite{FR,AMII} that  the Poisson structure on the phase space has a natural description in terms of Poisson--Lie group and coboundary Lie bialgebra structures associated with the isometry groups. The admissible classical $r$-matrices are characterised by the condition  that their symmetric component coincides with the element $t=P_a\oo J^a+J_a\oo P^a$ associated with the pairing \eqref{pair} or, equivalently, that their anti-symmetric component
solve the modified classical Yang--Baxter equation (YBE) \cite{FR}
\begin{align}\label{condition}
[[r,r]]=-\Omega\quad\text{with}\quad\Omega=[[t,t]],\quad t=J_a\oo P^a+P_a\oo J^a.
\end{align}
Although this does not define the classical $r$-matrices uniquely, there are
 strong indications that the relevant Lie bialgebra structures are the ones associated to Drinfel'd doubles \cite{BNR, we2, cm2, Woronowicz}. In this context, the associated quantum groups arise naturally as symmetries of the quantum theory and have a clear physical interpretation. The coproduct determines the composition of observables for multi-particle models as well as the implementation of constraints, while the antipode describes  anti-particles.  The universal $R$-matrix governs the exchange of particles through braid group symmetries and the ribbon element  the quantum action of the pure mapping class group.


\section{Quantum deformations of  $\rsl(2,\RR)$}

To relate the spacetimes and symmetry algebras of 3d gravity to quantum deformations,
we consider the real Lie algebra  $\rsl(2,\RR)\simeq \mathfrak{so}(2,1)$ with Lie bracket and star structure given by
\begin{align}
\label{liealg}
&[J_3,J_\pm]=\pm 2 J_ \pm  , \qquad [J_+,J_-]= J_3,\\
&J_3^*=-J_3 , \qquad J_\pm^*=-J_\pm.
\label{starstruct}
\end{align}
The universal enveloping algebra of $\rsl(2,\RR)$ can be endowed with a    non-deformed Hopf structure~\cite{CP}  with coproduct $\Delta_{(0)}:  \rsl(2,\RR)\to \rsl(2,\RR)\otimes \rsl(2,\RR)$,
\be
\Delta_{(0)}(J_i)=J_i\otimes 1 + 1\otimes J_i ,
\qquad i=3,\pm,
\label{cop}
\ee
which corresponds to the usual ``composition rule" for observables in the two particle case.

Up to equivalence, there are only {\em two} possible quantum ({i.e.}\ Hopf algebra) deformations for     $\rsl(2,\RR)$. The first one is the so-called
{\em standard or Drinfel'd--Jimbo deformation}, which was introduced in~\cite{Drinfelda,Jimbo} and reads:
\begin{align}
&[J_3,J_\pm]=\pm 2 J_\pm ,&    &[J_+ ,J_-]=\frac{\sinh (\m J_3)}{\m},
\label{brackets}\\ 
&\Delta( J_3) =J_3 \otimes 1 + 1\otimes J_3 ,&
  &\Delta (J_\pm) =J_\pm \otimes {\rm e}^{\frac \m 2 J_3} + {\rm e}^{-\frac \m 2 J_3} \otimes J_\pm  .
\label{cops} 
\end{align}
In the following, we denote it by  $\rsl_\m(2,\RR)$, where initially $\m$ is a {\em real} deformation parameter (usually  written in terms of  $q={\rm e}^\m$). In the limit  $\m\to 0$ (or $q\to 1$)  we recover (\ref{liealg}) and (\ref{cop}).

The expansion of  the deformed coproduct $\Delta$ as a formal power series in the parameter $\m$
\be
\Delta=\sum_{k=0}^{\infty} \Delta_{(k)}=\sum_{k=0}^{\infty} \m^k
\delta_{(k)}\,, 
\nonumber
\ee 
allows one to characterise quantum deformations of $\rsl(2,\RR)$ by the underlying Lie bialgebra structures. These are given by the  Lie algebra $\rsl(2,\RR)$ \eqref{liealg} together with the cocommutator $\delta: \rsl(2,\RR)\rightarrow \rsl(2,\RR)\oo \rsl(2,\RR)$ defined by the first-order
deformation of the  coproduct:
$$\delta =\m\cdot (\delta_{(1)} - \sigma\circ
\delta_{(1)}),
$$
where $\sigma(J_i\otimes J_j)=J_j\otimes
J_i$ is the flip map. For the deformation \eqref{cops} the cocommutator
reads \be
\delta(J_3)=0,\qquad  \delta(J_\pm)=\m J_\pm\wedge J_3 .
\label{cocostandard}
\ee
The associated Lie bialgebra structure is {\em coboundary}; the cocommutator is of the form
\begin{align}
&\delta(J_i)=[J_i\otimes 1 + 1\otimes J_i,r_\m],\qquad i=3,\pm\,,
\label{coco}\\
&r_\m=\m J_+\wedge J_- = \m(J_+\otimes J_- - J_-\otimes J_+)\label{rstandard},
\end{align}
where $r_\m$
is a {\em classical r-matrix}, i.e.~a constant solution of the {\em modified} classical YBE $$[[r_\m,r_\m]]=-\m^2\Theta,\qquad\Theta=J_3\otimes J_-\otimes J_+-J_3\otimes J_+\otimes J_-+ \text{cyclic permutations}.$$

We recall that the quantum algebra $\rsl_\m(2,\RR)\simeq \mathfrak{so}_\m(2,1)$ is the rank-one case within the 
series  of the   quantum   $\mathfrak{so}_\m(p,q)$ algebras of  Drinfel'd--Jimbo type. Under quantum contractions~\cite{CGST2,LBC}, these quantum algebras have provided the well-known $\kappa$-Poincar\'e algebra~\cite{Lukierskia, Giller, Lukierskib, Maslanka}  as well as its associated $\kappa$-Minkowski spacetime~\cite{Majid:1994cy, Zak, LukNR} in which $\kappa=1/\m$.   In this framework, the  deformation parameter $\kappa$  has been interpreted as a second
observer-independent fundamental scale in addition to  the speed of light $c$, which would be related with the Planck length and, presumably, with the cosmological constant~\cite{amel}.


A second {\em nonstandard  or Jordanian  twist deformation}  for  $\rsl(2,\RR)$   was introduced in~\cite{Ohn}. 
We denote  it by  $\rsl_z(2,\RR)$ where $z$ is   a  {\em real} deformation parameter ($q={\rm e}^z$). Its
 commutation rules and coproduct    read
\begin{align}
&[J_3,J_+ ] =  \frac{4\sinh (\frac z 2 J_+ )}z , 
\quad[J_3,J_-] = -J_-\cosh (  z J_+/2 )  - \cosh (  z J_+/2 ) J_- ,\quad [J_+ ,J_-] =  J_3 ,
\nonumber \\
&\Delta(J_+)     =  J_+ \otimes 1 +  1 \otimes J_+ , \qquad\ 
\Delta(J_l) =  J_l \otimes
{\rm e}^{\frac z2 J_+ }    +  {\rm e}^{- \frac z2 J_+ } \otimes J_l  , \qquad l=3,-.
\label{cosu2n} \end{align}
The  limit $z\to 0$ again  reproduces the non-deformed Hopf algebra structure of $\rsl(2,\RR)$, and the associated Lie bialgebra structure is  coboundary with classical $r$-matrix and  cocommutator
\be
r_z=\frac z2 J_3\wedge J_+ ,\qquad \delta(J_+)=0,\qquad \delta(J_l)=z J_l\wedge J_+,\qquad l=3,-.
\label{rnstandard}
\ee
Note that for this deformation, the classical $r$-matrix $r_z$ is  a constant solution of the {\em unmodified} classical YBE: $[[r_z,r_z]]=0$.
  
This quantum deformation has been used in the construction of higher dimensional      nonstandard quantum $\mathfrak{so}(p,q)$ algebras~\cite{beyond, Lukierskicc, Lukierskicd, vulpi, aizawa}, which have an interpetation as quantum deformations of conformal symmetries. In this context, the deformation parameter $z$ plays the role of the lattice step  on uniform discretisations of the Minkowski space.
The nonstandard deformation of $\rsl(2,\RR)$ also defines the  so called ``null-plane'' quantum Poincar\'e algebra~\cite{Nulla} which gave rise to  non-commutative Minkowskian spacetimes~\cite{Nullb,Brunoc} different from the $\kappa$-Minkowski one. For recent applications of twist deformations  in the construction of non-commutative Minkowskian spacetimes, see~\cite{Kim,Borowiec} and the references therein. Finally, this twist deformation
 has  also  been used to obtain ``deformed"   AdS and dS spacetimes,   understood as spaces endowed with a non-constant curvature governed by the deformation parameter $z$~\cite{plbcurvature,IJTP}.


Although quantum deformations generally do not admit superpositions,  it turns out that the
standard and nonstandard deformations  introduced above can be superposed, giving rise to the so-called {\em ``hybrid" deformation} of  $\rsl(2,\RR)$~\cite{gl2}, which we denote by $\rsl_{\m,z}(2,\RR)$ in the following. 
In this case the two-parametric classical $r$-matrix
\be
r=r_\eta+r_z=\m J_+\wedge J_- + \frac z2 J_3\wedge J_+,
\label{rhybrid}
\ee
is of standard type, i.e.~a solution of the {\em modified} classical  YBE: $[[r,r]]=-\m^2\Theta$. The associated cocommutator, given by (\ref{coco}), is the  sum of (\ref{cocostandard}) and (\ref{rnstandard}):
\begin{align}
\label{cocomm}
\delta(J_+)=\m J_+\wedge J_3, \qquad \delta(J_3)=z J_3\wedge J_+, \qquad \delta(J_-)=\m J_-\wedge J_3+z J_-\wedge J_+.
\end{align}
The  full quantum Hopf structure of $\rsl_{\m,z}(2,\RR)$ is rather involved and can be found in~\cite{gl2}. For our purposes, it is sufficient to consider its lowest order terms, i.e.~the Lie bialgebra structure defined by \eqref{liealg} and  \eqref{cocomm}.

From a purely mathematical viewpoint, the fact that the classical $r$-matrix (\ref{rhybrid}) is of a standard type makes this deformation equivalent to the Drinfel'd--Jimbo one through an appropriate change of basis which was achieved in~\cite{dobrev}. Consequently,  the deformation parameter $z$ 
would be viewed as non-essential. 
However, we will show in the following that in the context of 3d gravity both parameters play essential roles and have a clear physical interpretation.


\section{3d AdS gravity from the  ``hybrid'' Drinfel'd double}

As explained previously,  each quantum deformation of $\rsl(2,\RR)$ gives rise to a unique coboundary Lie bialgebra structure $(\rsl(2,\RR), \delta)$ characterised by a classical $r$-matrix. Conversely, each coboundary Lie bialgebra associated with $\rsl(2,\RR)$ gives rise to a Drinfel'd double Lie algebra~\cite{gomez,ab1}. In this section, we construct the Drinfel'd double for the ``hybrid" deformation $\rsl_{\m,z}(2,\RR)$, for real deformation parameters, and show that this generates the AdS symmetry algebra of 3d gravity shown in Table 1.

We consider the ``hybrid" deformation  $\rsl_{\m,z}(2,\RR)$ and denote by 
  $A^k_{ij}$ the structure constants of the Lie algebra $\rsl(2,\RR)$ (\ref{liealg})  and by $B^{ij}_k$ the structure constants of the cocommutator \eqref{cocomm} with respect to the basis $\{J_3,J_\pm\}$ 
  \begin{align}
\label{}[J_i,J_j]=A_{ij}^kJ_k ,\qquad \delta(J_i)=B_i^{jk}J_j\oo J_k ,\qquad i,j,k=3,\pm.
\end{align}
As a Lie algebra, the  classical Drinfel'd double $\calD_{\m,z}(\rsl(2,\RR),\delta)$ is the six-dimensional Lie algebra
spanned by the basis $\{J_i\}_{i=3,\pm}$  and its dual basis $\{j^i\}_{i=3,\pm}$ with Lie brackets 
\begin{align}
[J_i,J_j]=A_{ij}^kJ_k,\qquad [j^i,j^j]=B^{ij}_k j^k,\qquad [J_i, j^j]=B_i^{jk}J_k-A_{ik}^j j^k .
 \end{align}
The full set of Lie brackets defining $\calD_{\m,z}(\rsl(2,\RR),\delta)$ thus consists of  the brackets \eqref{liealg} of $\rsl(2,\RR)$,  the brackets of its dual Lie algebra induced by the cocommutator \eqref{cocomm}
 \begin{align}
\label{dualb}
&[j^3,j^+]=-\m j^++zj^3,\qquad[j^3,j^-]=-\m j^-,\qquad [j^+,j^-]=-zj^-,
\end{align}
and the ``crossed" or ``mixed" Lie brackets  
  \be 
\begin{array}{lll}
[J_3,j^3]=zJ_+  ,&\quad  [J_3,j^+]=-zJ_3-2j^+  ,&\quad [J_3,j^-]=2j^- ,\\[2pt]
[J_+, j^3]=- \m  J_+- j^-  ,&\quad [J_+,j^+]=\m J_3+2j^3 ,&\quad [J_+,j^-]=0 ,\\[2pt]
[J_-,j^3]=-\m J_-+ j^+ ,&\quad [J_-,j^+]=-zJ_-  ,&\quad [J_-,j^-]=\m J_3+zJ_+-2j^3 .
   \end{array}
\label{mixb}
\ee 
The cocommutator of $\calD_{\m,z}(\rsl(2,\RR),\delta)$ is obtained via \eqref{coco} from its classical $r$-matrix
\be
r=\sum_{i=3,\pm}{j^i\otimes J_i},
\label{rdouble}
\ee
which  induces the pairing between  the basis $\{J_i\}_{i=3,\pm}$ and the dual basis $\{j^i\}_{i=3,\pm}$
\begin{align}
&\langle J_i,j^k\rangle=\langle j^k,J_i\rangle=\delta_i^k , \qquad \langle J_i,J_k\rangle=\langle j^i,j^k\rangle=0 , \qquad i,k=3,\pm.\label{pairpm}
\end{align}
If both deformation parameters are real, $\calD_{\m,z}(\rsl(2,\RR),\delta)$ inherits   a star structure from the star structure \eqref{starstruct}  of $\rsl(2,\RR)$
 \be
\label{star} 
J_3^*=-J_3, \qquad J_\pm^*=-J_\pm ,\qquad
 j^{3*}=-j^3 ,\qquad j^{\pm*}=-j^\pm.
\ee
The essential step in relating the ``hybrid" quantum deformation of $\rsl(2,\RR)$ to the spacetimes, symmetry algebras and quantum groups of 3d gravity~\cite{amel, cm1, cm2} is the introduction of a new basis of $\calD_{\m,z}(\rsl(2,\RR),\delta)$, in the following referred to as {\em Chern--Simons  basis}. This basis consists of generators
$J_a, P_a$, $a=0,1,2$, that are related to the generators of the hybrid Drinfel'd double  $\calD_{\m,z}(\rsl(2,\RR),\delta)$ as follows
\begin{align}
&J_0=\frac 1 2 (J_+\!-\!J_-) ,& 
&J_1=\frac z 2 J_3 , &  &J_2=\frac z 2 (J_+\!\!+\!J_-)  , \label{csbasis}\\
&P_0\!=\!  \m  (J_+\!\!+\!J_-)\!-\! \frac {z} 2 J_3\!+\!j^-\!\!\!-\!j^+ ,& 
&P_1\!=\! -z^2 \! J_+\!\!+\!2 z j^3  ,&  &P_2\!=\!   \m  z (J_+\!\!-\!J_-)\!+\!\frac {z^2} 2  J_3\!+\! z(j^+\!\!+\!j^- ) .\nonumber
\end{align}
Using expressions  (\ref{liealg}), (\ref{dualb}) and (\ref{mixb}) for the Lie brackets of  $\calD_{\m,z}(\rsl(2,\RR),\delta)$, we find that the Lie brackets in the CS basis take the form
 \be 
\begin{array}{lll}
 [J_0,J_1]=-J_2, & \quad [J_0,J_2]=J_1,  & \quad  [J_1, J_2]=z^2 J_0,  \\[2pt]
[J_0,P_0]=0 ,& \quad [J_0,P_1]=- P_2 ,& \quad [J_0, P_2]=P_1,\\[2pt]
[J_1,P_0]=P_2 ,& \quad [J_1,P_1]=0 ,& \quad [J_1, P_2]=z^2 P_0,\\[2pt]
[J_2,P_0]=-P_1 ,& \quad[J_2,P_1]=- z^2 P_0 ,& \quad[J_2, P_2]=0,\\[2pt]
[P_0,P_1]=-  {4\m^2}  J_2, & \quad[P_0, P_2]={4\m^2} J_1  ,& \quad  [P_1,P_2]= {4\m^2 z^2}  J_0   .
 \end{array}
\label{jj}
\ee 
Provided that   the deformation parameters $\m,z$ are {\em non-zero real} numbers,  we have
\be
  \calD_{\m,z}(\rsl(2,\RR),\delta)\simeq  \mathfrak{so}(2,2) .
 \label{ka}
\ee
The deformation parameters thus enter the Lie bracket \eqref{jj} in the CS basis as  structure constants $4\m^2$ and $z^2$ which can be set equal  to $+1$ by rescaling the generators  as
\begin{align}
4\m^2\to 1:\  P_a\to \frac{1}{2\m} P_a \quad (a=0,1,2) ;\qquad 
z^2\to 1: &\  P_b\to \frac{1}{z} P_b,\  J_b\to \frac{1}{z} J_b \quad (b=1,2).
\label{kab}
\end{align}
If the elements of the CS basis  $J_0,J_b,P_0,P_b$, $b=1,2$, are interpreted, in this order, as the generators of rotations, boosts, time translations and spatial translations, then $ \mathfrak{so}(2,2)$   can be  identified with  symmetry algebra of the 3d  AdS  space, in which $J_0,J_1,J_2$ span the Lorentz subalgebra $ \mathfrak{so}(2,1)$. The AdS
spacetime is then obtained as  the homogenous space $\mathbf{AdS}^{2+1}\!\!=\!{SO}(2,2)/  {SO}(2,1)$ where   $J_1$ and $J_2$ are the generators of inertial transformations along the 2 and 1 directions.

Surprisingly enough, this  result  gives rise to a direct identification between the deformation parameters of the hybrid deformation $\rsl_{\m,z}(2,\RR)$ and the physical parameters of the 3d gravity: the  cosmological constant $\Lambda=-\lam$  (\ref{lambdadef}) and the speed of light $c$, which are given by
 \be
\lam = 4\m^2,\qquad c^2=1/z^2.
 \label{kb}
 \ee
 In other words, $\m$ determines the cosmological constant (and hence  the curvature), while $z$ characterises the signature of the metric $g$  as
 \be
   g=\mbox {diag}(-1, z^2,z^2).
\label{metric}
 \ee
The other two essential ingredients in the CS formulation of 3d-gravity are the Ad-invariant symmetric bilinear form \eqref{pair} on the symmetry algebra and the star structure \eqref{starcs}. Using the relations \eqref{csbasis} between the  original Drinfel'd basis and the CS basis, 
we find that the star structure 
\eqref{star}  induces  the star structure \eqref{starcs}.
Moreover, up to a rescaling with $z$, which sets the speed of light to one, the
 pairing \eqref{pairpm} agrees with the pairing \eqref{pair} in the CS action 
\be
\label{jppair}
 \langle J_0, P_0\rangle=-1 ,\qquad \langle J_1, P_1\rangle=z^2 ,\qquad \langle J_2, P_2\rangle=z^2.
\ee
The hybrid deformation $\rsl_{\m,z}(2,\RR)$ with non-zero real  parameters $\m,z$ thus reproduces all relevant structures that enter into the CS formulation of Lorentzian 3d gravity with negative cosmological constant: the Lie bracket \eqref{brack}, the star structure~\eqref{starcs} and the pairing~\eqref{pair}.

It is instructive to express the classical $r$-matrix \eqref{rdouble} of $\calD_{\m,z}(\rsl(2,\RR),\delta)$  in the CS basis:
 \be
r=\frac 1 z \left( z J_0\wedge J_1 +  {2}\eta\, J_2\wedge J_0+   J_2\wedge J_1 \right) + \frac{1}{z^2}\left( P_1\otimes J_1+ P_2\otimes J_2 -  z^2 P_0\otimes J_0\right).
\label{rCS}
\ee
The resulting $r$-matrix consists of two terms: the first is spanned by the Lorentz subalgebra while the second one is related to the angular momentum  or Pauli--Lubanski invariant.
This classical $r$-matrix is a solution of the modified classical YBE  \eqref{condition}, which implies that the associated  quantum group symmetries are compatible with 3d gravity.


\section{Spacetimes and symmetry algebras of 3d gravity}

The results of the last section demonstrate that the hybrid Drinfel'd double naturally gives rise to all  data that defines 3d AdS gravity and at the same time  provides a physical interpretation of both deformation parameters.  We will now generalise this result to other signatures and values of the cosmological constant through analytic continuation and contractions.

\begin{table}[t] {\footnotesize
 \noindent
\caption{{\small The nine homogeneous 3d spaces obtained from the hybrid Drinfel'd double  according to the possible values of the deformation parameters $\m,z$. The signature of the pairing together with the star structure of the Drinfel'd basis for $z\ne 0$ are also displayed.
 }}
\label{table1}
\medskip
\noindent\hfill
$$
\begin{array}{lll}
\hline
\\ [-6pt]
\multispan{3}{\qquad \qquad\qquad \qquad \qquad \qquad \qquad \qquad \mbox{Riemannian spaces}}\\[4pt]
\hline
\\ [-6pt]
\mbox {$\bullet$   Three-sphere} &\quad\mbox {$\bullet$  Euclidean  space
 }&\quad\mbox {$\bullet$ Hyperbolic space   } \\[2pt] 
 {\bf S}^3=SO(4)/SO(3)  &\quad  {\bf E}^3=  ISO(3)/SO(3)
 &\quad     {\bf H}^3=   SO(3,1)/SO(3) \\[2pt] 
  \m\in\RR^\ast,\ z\in {\rm i}\RR^\ast
 &\quad  \m=0 ,\ z\in {\rm i}\RR^\ast&\quad  \m\in{\rm i}\RR^\ast,\ z\in {\rm i}\RR^\ast \\[2pt]
\Lambda=\lam>0, \   c\in {\rm i}\RR^\ast
 &\quad  \Lambda=\lam=0, \   c\in {\rm i}\RR^\ast &\quad \Lambda= \lam<0, \   c \in {\rm i}\RR^\ast\\[2pt]
\pairing=(-1,-1,-1)
 &\quad  \pairing =(-1,-1,-1) &\quad  \pairing =(-1,-1,-1)\\[2pt]
 J_3^\ast=J_3,\ J_\pm^\ast =J_\mp  &\quad
 J_3^\ast=J_3,\ J_\pm^\ast =J_\mp  &\quad
 J_3^\ast=J_3,\ J_\pm^\ast =J_\mp   \\[2pt]
  j^{3 \ast}=j^{3}-\frac z2 (J_+ +J_-) &\quad
 j^{3 \ast}=j^{3}-\frac z2 (J_+ +J_-)    &\quad
 j^{3 \ast}=j^{3}-\frac z2 (J_+ +J_-) \\[2pt]
 j^{\pm\ast}=j^{\mp}+\frac z2 J_3   \pm 2 \m J_\pm &\quad
 j^{\pm\ast}=j^{\mp}+\frac z2 J_3   &\quad
 j^{\pm\ast}=j^{\mp}+\frac z2 J_3   
  \\[6pt]
   \hline
\\ [-6pt]
\multispan{3}{\qquad \qquad\qquad \qquad \qquad \qquad \qquad \qquad \mbox{Newtonian spaces}}\\[4pt]
\hline
\\ [-6pt]
\mbox {$\bullet$  Oscillating NH space}&\quad\mbox {$\bullet$ Galilean space   }&\quad\mbox {$\bullet$
Expanding NH space }\\[2pt] 
\mbox {${\bf NH}^{2+1}_+={\rm  NH}_+/ISO(2)  $}&\quad\mbox {${\bf
G}^{2+1}=IISO(2)/ISO(2) $}&\quad\mbox
{${\bf NH}^{2+1}_-=  {\rm  NH}_-/ISO(2) $}\\[2pt] 
    \m\in\RR^\ast,\ z=0
 &\quad  \m=0 ,\ z=0&\quad  \m\in{\rm i}\RR^\ast,\ z=0 \\[2pt]
\lam>0, \   c= \infty
 &\quad  \lam=0, \    c= \infty&\quad \lam<0, \    c= \infty\\[2pt]
\pairing=(-1,0,0)
 &\quad  \pairing =(-1,0,0)&\quad  \pairing =(-1,0,0)
  \\[6pt]
      \hline
\\ [-6pt]
\multispan{3}{\qquad \qquad\qquad \qquad \qquad \qquad \qquad \qquad \mbox{Lorentzian spaces}}\\[4pt]
\hline
\\ [-6pt]
\mbox {$\bullet$ AdS  space  }&\quad\mbox {$\bullet$ Minkowski  space  }&\quad\mbox {$\bullet$ dS  space }\\[2pt] 
{\bf AdS}^{2+1}=  SO(2,2)/SO(2,1) &\quad  {\bf
M}^{2+1}=  ISO(2,1)/SO(2,1)
 &\quad    {\bf dS}^{2+1}=  SO(3,1)/SO(2,1)    \\[2pt] 
  \m\in\RR^\ast,\ z\in \RR^\ast
 &\quad  \m=0 ,\ z\in \RR^\ast&\quad  \m\in{\rm i}\RR^\ast,\ z\in \RR^\ast \\[2pt]
 \Lambda<0,\  \lam>0, \   c>0
 &\quad \Lambda= \lam=0, \   c>0 &\quad  \Lambda>0,\   \lam<0, \   c>0\\[2pt]
\pairing=(-1,+1,+1)
 &\quad  \pairing =(-1,+1,+1) &\quad  \pairing =(-1,+1,+1)
    \\[2pt]
J_3^\ast=-J_3,\ J_\pm^\ast =-J_\pm  &\quad
J_3^\ast=-J_3,\ J_\pm^\ast =-J_\pm   &\quad
J_3^\ast=-J_3,\ J_\pm^\ast =-J_\pm    \\[2pt]
  j^{3 \ast}=-j^{3} ,\  j^{\pm\ast}=-j^{\pm}&\quad
 j^{3 \ast}=-j^{3} ,\ j^{\pm\ast}=-j^{\pm}  &\quad
 j^{3 \ast}=-j^{3},\ j^{\pm\ast}=-j^{\pm}\pm 2\m J_\mp   
\\[6pt]
 \hline
\end{array}
$$
\hfill}
\end{table}

For this purpose we note that expressions \eqref{csbasis} and \eqref{jj} for the CS basis and the Lie brackets are well defined also for  imaginary values of $\m,z$.
If, additionally, we consider  the limits $\m,z\to 0$,  we obtain nine 3d homogenous spaces ${\bf X}_{\m,z}$, which are a subfamily of the Cayley--Klein spaces~\cite{IJTP}. 
They are given as the quotient of the Lie group associated with the Lie algebra (\ref{jj}) by
 the subgroup spanned by  the three generators $J_a$
 $$
{ \bf X}_{\m,z}=\langle \calD_{\m,z}(\rsl(2,\RR),\delta)\rangle/\langle J_0,J_1,J_2\rangle .
 $$
As in the AdS case, we find that  the Lie groups associated with the Drinfel'd double  via \eqref{jj} act as the isometry groups of these spaces. The parameters $\m,z$ define, respectively, their curvature $\lambda$ and the speed of light $c$ via \eqref{kb}.  Therefore, we recover the six spacetimes  of Table 1 with cosmological constant $\Lambda=\pm\lambda$ \eqref{lambdadef}  whenever  $z\ne0$. These nine spaces are presented in Table \ref{table1}, together with the corresponding values of $\Lambda,\lambda$ and $c$.

Note that the expression for the pairing \eqref{jppair} coincides with the one for 3d gravity \eqref{pair} for all non-zero values of $z$, i.e.~whenever a relevant 3d gravity model exists.
This pairing defines the signature
 $\pairing= 
{\rm signature}( -1,z^2,z^2)$ of the associated homogeneous spaces  in Table \ref{table1}.
However, in order to obtain the star structure \eqref{starcs} for the CS formulation of 3d gravity, we need to impose different star structures on the  
initial Drinfel'd basis $J_3,J_\pm,j^3,j^\pm$ which are listed in Table \ref{table1}  for $z\neq 0$.

The limits $\m\to 0$ and $z\to 0$ are  well defined for the Lie algebra (\ref{jj}) as well as the pairing (\ref{jppair}). They  can be understood, respectively,  as the ``flat" and ``non-relativistic"  In\"on\"u--Wigner contractions. Explicitly, if we start with a Lie algebra $\calD_{\m,z}(\rsl(2,\RR),\delta)$ with one of the two deformation parameters fixed to a non-zero value, then  the Lie algebra contractions are obtained  via  a rescaling of the CS basis together with the corresponding limit:
\begin{align}
&{\mbox {``Flat"\ contraction:}} &  &P_a\to\m  P_a, &  &a=0,1,2, &  &\m\to 0.\nonumber\\
 &{\mbox{``Non-relativistic"\ contraction:}} &  &P_b\to z P_b,\;\; J_b\to z J_b ,& &b=1,2, &  &z \to0.
\nonumber
\end{align}
 In these transformations the parameters $\m$ and $z$ have a proper interpretation as {\em contraction parameters}, and the rescaling of the generators is the inverse of   (\ref{kab}). Note, however, that in the limit $z\to 0$ the basis transformation  (\ref{csbasis}) becomes 
singular and the $r$-matrix \eqref{rdouble} diverges, thus precluding the use of the initial Drinfel'd basis for the three spaces with $z=0$.  
Nevertheless, this limit can be performed for the $r$-matrix (\ref{rCS}) if it is combined with a rescaling $r\to z^2 r$. The resulting $r$-matrix, 
$r= P_1\otimes J_1+ P_2\otimes J_2  $,
is a Reshetikhin twist as all the generators contained in it commute for $z\rightarrow 0$.

To summarise,  the quantum algebra  $\rsl_{\m,z}(2,\RR)$ gives rise to nine   homogeneous spaces:
\begin{itemize}

\item For $z\in {\rm i} \RR^\ast$, the parameter $z$ can be set to $i$ via the rescaling \eqref{kab}.
 We obtain the  three  classical {\em Riemannian 3d spaces} of constant curvature: the three-sphere,  3d hyperbolic space   and 3d Euclidean space. In these cases $\Lambda=\lambda =4\m^2=\pm 1/R^2$, where $R$ is the
radius of the space  ($R\to \infty$ for ${\mathbf E}^3$). 
These are the three relevant models for Euclidean 3d gravity given in Table 1. Note that the limit $\m\to 0$ corresponds to the  well-known  {flat contraction}  $\mathfrak{so}(4)\to \mathfrak{iso}(3)
\leftarrow \mathfrak{so}(3,1)$. 

\item For  $z\in  \RR^\ast$, the parameter $z$ can be set to $1$ through \eqref{kab}. This yields  the three standard {\em Lorentzian 3d spacetimes} of constant curvature:   3d  AdS,  dS and Minkowski space. Now  $\Lambda=-\lambda$ with $\lambda=4\m^2=\pm 1/\tau^2$, where $\tau$ is the (time) universe radius  ($\tau\to \infty$ for  ${\bf M}^{2+1}$), so we recover the 
 three relevant models for Lorentzian 3d gravity given in Table 1.
  The limit $\eta\rightarrow 0$ yields   the contraction $\mathfrak{so}(2,2)\to \mathfrak{iso}(2,1)\leftarrow \mathfrak{so}(3,1)$.

\item The limit $z=0$ ($c\to \infty)$  gives rise to three non-relativistic or {\em Newtonian spacetimes} which cover the two   Newton--Hooke (NH) curved spacetimes~\cite{IJTP, BLL, Aldrovandi} and the flat Galilean one. As both the metric \eqref{metric} and the pairing \eqref{jppair}  become degenerate, these 
models  do not describe standard 3d gravity in which the metric is required to be non-degenerate and of either Euclidean or Lorentzian signature. However, these spaces are of interest as they arise in the non-relativistic or Galilean limit of the theory \cite{bernd}.
The associated isometry groups are semidirect  products
$$
{\rm NH}_+=T_4\rtimes (SO(2)\otimes SO(2)) ,\qquad {\rm NH}_-= T_4\rtimes(SO(1,1)\otimes SO(2)),
$$
where $T_4$ is the four-dimensional abelian Lie algebra spanned by $\{P_b,J_b\}$, $b=1,2$.


\end{itemize}

To conclude, we remark that
the quantum algebra $\rsl_{\m,z}(2,\RR)$ could    be used as the cornerstone for a unified construction of 3d  doubly special relativity theories with a non-zero cosmological constant and with either Lorentzian or Euclidean
signature. Also, we note that the symmetry algebras of 3d gravity obtained from the hybrid deformation $\rsl_{\m,z}(2,\RR)$ coincide with the Lie algebras that arise as the conformal symmetries of  2d constant curvature spacetimes in \cite{conf}. Since the curvature and signature parameters of the latter correspond, respectively, to the  signature and curvature parameters of the 3d spacetimes, it would be interesting to explore this duality further and to clarify its interpretation in the context of 3d gravity.


{\small

\section*{Acknowledgements}

This work was partially supported by the Spanish Ministerio de   Ciencia e Innovaci\'on    under grant     MTM2007-67389 (with EU-FEDER support) and by  Junta de Castilla y
Le\'on  (Project GR224). C.M. thanks the University of Burgos for hospitality during a research visit in January 2009. Her work was supported by the Marie Curie Intra-European Fellowship PIEF-GA-2008-220480 (until March 2009) and by the DFG Emmy-Noether fellowship  ME 3425/1-1 (from April 2009).
}



\end{document}